\documentclass[aps,pre,twocolumn]{revtex4}
\usepackage{amsmath,bm,epsfig}

\def\Fbox#1{\vskip1ex\hbox to 8.5cm{\hfil\fboxsep0.3cm\fbox{%
  \parbox{8.0cm}{#1}}\hfil}\vskip1ex\noindent}  

\newcommand{\B}[1]{{\bm{#1}}}
\let \= \equiv \let\*\cdot \let\~\widetilde \let\^\widehat \let\-\overline

\newcommand{\ud}{\mathrm{d}}
\def\<{\left\langle}    \def\>{\right\rangle}
\def\({\left(}          \def\){\right)}
 \def \[ {\left [} \def \] {\right ]}

\begin{document}
\title{Theory of Specific Heat in Glass Forming Systems}
\author{ H.G.E. Hentschel$^*$, Valery Ilyin, Itamar Procaccia and Nurith Schupper}
\affiliation{Department of Chemical Physics, The Weizmann
Institute of Science, Rehovot 76100, Israel \\
$^*$ Dept of Physics, Emory University, Atlanta Ga 30322. }
\date{\today}
\begin{abstract}
Experimental measurements of the specific heat in glass-forming systems are obtained from the linear response to either slow cooling (or heating) or to oscillatory perturbations with a given frequency about
a constant temperature. The latter method gives rise to a complex specific heat with the constraint that the zero frequency limit of the real part should be identified with thermodynamic measurements. Such measurements reveal anomalies in the temperature dependence of the specific heat, including the so called ``specific heat peak" in the vicinity of the glass transition. The aim of this paper is to provide theoretical explanations of these anomalies in general and a quantitative theory in the case of a
simple model of glass-formation.  We first present new simulation results for the specific heat
in a classical model of a binary mixture glass-former.  We show that in addition to the formerly observed specific heat peak there is a second peak at lower temperatures which was not observable
in earlier simulations. Second, we present a general relation between the specific heat, a caloric quantity, and the bulk modulus of the material, a mechanical quantity, and thus offer a smooth connection between the liquid and amorphous solid states. The central result of this paper is a connection between the micro-melting of clusters in the system and the appearance of specific heat peaks; we explain the appearance of two peaks by the micro-melting of two types of clusters. We relate the two peaks to changes in the bulk and shear moduli.  We propose that the phenomenon of glass-formation is accompanied by a fast change in the bulk and the shear moduli, but these fast changes occur in different ranges of the temperature. Lastly,  we demonstrate how to construct a theory of the frequency dependent complex specific heat, expected from heterogeneous clustering in the liquid state of glass formers. A specific example is provided in the context of our model for the dynamics of glycerol. We show that the frequency dependence is determined by the same $\alpha$-relaxation mechanism that operates when measuring the viscosity or the dielectric relaxation spectrum. The theoretical frequency dependent specific heat agrees well with experimental measurements on glycerol. We conclude the paper by stating that there is nothing universal about
the temperature dependence of the specific heat in glass formers - unfortunately one needs to understand 
each case by itself.

\end{abstract}
 \maketitle
 
 \section{Introduction}
 
 The traditional measurements of the specific heat $C_V$ at constant volume or $C_P$ at constant
 pressure involve cooling (or heating) the sample at a constant rate \cite{68SS,76AS}. When applied to glass-forming
 systems, this approach has an inherent difficulty. Since glass-forming system tend to relax to equilibrium slower and slower as the temperature is lowered, at some point the `constant rate' of cooling becomes too high for the system to respond to, and then the system does not reach equilibrium. Typically the specific heat then drops abruptly, giving rise to the ``specific heat peak" at some temperature which is 
 sometimes identified as the glass transition temperature $T_g$. Needless to say, such a definition of
 transition temperature is less than compelling, since it clearly depends on the rate of cooling and is not inherent to the system properties.
 
 In an attempt to overcome this difficulty Birge and Nagel \cite{85BN} introduced "specific heat spectroscopy". In this technique one keeps the sample close to a temperature $T$ at all times, but perturbs it periodically
 with a small-amplitude oscillation of frequency $\omega$. Linear response theory then relates
 the amount of heat exchanged at that frequency, $\delta Q(\omega)$ to the oscillatory temperature 
 perturbation $\delta T(\omega)$ via the relation 
 \begin{equation}
\delta  Q(\omega) = C(\omega) \delta T(\omega)
 \end{equation}
 where $C(\omega)$ is the frequency dependent specific heat that can be measured at
 either constant volume or constant pressure. In order to find the thermodynamic specific heat one
 needs to extrapolate data to the $\omega\to 0$ limit. Whether or not this extrapolation overcomes the above-mentioned worry of sufficient relaxation time is an issue that has not been fully clarified in the literature. 
 
\begin{figure}
\centering
~\hskip -1.4 cm
\includegraphics[width=0.65\textwidth]{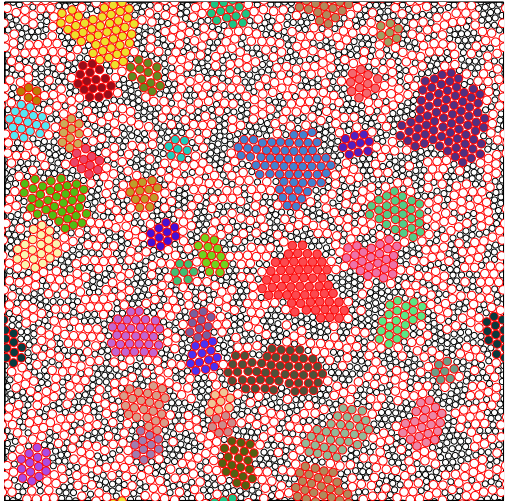}
\caption{(Color online). A snapshot of the system at $T=0.44$. In colours we highlight the clusters of large particles in  local hexagonal order. The colours have no meaning.}
\label{clusters}
\end{figure}

 In this paper we concern ourselves with the theoretical calculation of the specific heat in glass-forming
 systems and in the relation of the specific heat to other material properties. To this aim we focus
 on one simulational example (a binary mixture of point particles interacting via an $r^{-12}$ repulsive potential) and one experimental example (glycerol). In the context of the first example we present results of new simulations that exhibit two distinct peaks in the curve of the specific heat vs. the temperature. We present for this example various theoretical results, culminating with a new
 scenario to explain the specific heat peaks, i.e. the micro-melting of clusters. We believe that this is the central point of the present paper. To understand the nature of the specific heat anomalies one must understand the physics that is behind the glassy behavior of this model in general and the existence of the two specific heat peaks in particular. When the temperature is lowered at a fixed pressure this system \cite{08LP} (as well as many other glass-formers \cite{06ST,09LPR,GS82,93Cha,98KT,00TKV}  tends to form micro-clusters of local order. In the present case 
large particles form long-lived patches of hexagonal ordering first (starting at about $T=0.5$, and at lower temperatures (around $T=0.1$) also the small particles form long-lived hexagonal clusters. The clusters are not that huge, with at most O(100) particles, (cf. Fig. \ref{clusters}), depending on the temperature and the aging time. But we have shown that the long time properties
of correlation functions are entirely carried by the micro-clusters \cite{08LP}. Below we will refer to the micro-clusters as curds and the liquid phase as whey.
We will argue that the specific heat responds to the micro-melting of the clusters - those of small particles at the lowest temperatures and those of the larger particles at higher temperatures. The large increase in
the number of degrees of freedom when a particle leaves a crystalline cluster and joins the liquid background is the basic reason for the increase in entropy that is seen as a specific heat peak.
 
 In the context of the second example we show that 
the calculation of the frequency dependent  specific heat is easy when we have
a reasonable model of the glassy relaxation. Having
 such a model for glycerol \cite{08HP}, we demonstrate in section \ref{glycerol} that the information gained from the frequency dependent specific heat is very similar to that learned from other linear
 response functions like broad-band dielectric spectroscopy. We will be able therefore to present
 spectra of the frequency dependent specific heat in close correspondence with experiments. 
  
 The specific heat has interesting relations with the mechanical moduli of the material, and
 we present relations (which pertain to any system with an $r^{-n}$ potential) to the bulk and shear moduli. As a result of our thinking we conclude that the bulk and shear moduli change rapidly in the temperature range of the
two distinct specific heat peaks mentioned above. The relation to the bulk modulus is explicit,
and is shown rigorously in Subsect.~\ref{mechanics}. The relation to the shear modulus is less explicit, simply because one does not have an equation of state with strain (a quantity that is ill defined in the context of glasses).  Nevertheless we present a conjecture that the bulk and shear moduli in generic glasses may change rapidly at two different temperature ranges. 

 The structure of the paper is as follows: In Sect. \ref{Binary} we present the model glass consisting of a binary mixture of point particles interacting via soft potentials, and discuss its thermodynamic properties. We derive
 exact equations for its specific heat at constant volume, which are correct at all temperatures through
 the glass transition. We compare these results to molecular dynamics simulations in which special care had been taken to equilibrate the system, summarizing a computational effort of about two years. The main conclusion of this section is that the details of the interaction potential are crucial in determining
 what the specific heat does in the vicinity of the glass transition, and there is nothing universal about it. For a simple enough potential we can derive a theory that is in excellent agreement with simulations
 up to the first specific heat peak. To explain both peaks we must present a theory that takes into
 account explicitly the tendency of the system to form micro-clusters \cite{08LP}. The state of the system
 then becomes like curds of local crystalline order embedded in a whey of disordered fluid. It is the freezing or melting of these curds that account quantitatively for the specific heat peaks, as is shown in Subsect. \ref{physics}. Below we use interchangeably the words `clusters' and `curds'.
 In Sect. \ref{glycerol} we turn to discussing the frequency-dependent complex specific heat. To construct a theory of this object one needs a model of the dynamics of the system under study, be it glycerol or any other material. We demonstrate, using our dynamical model of glycerol \cite{08HP}, how this measurement is equivalent in terms of
 its dynamical contents to any other linear response to an oscillatory perturbation. We present theoretical spectra and show satisfactory agreement with the
experiments.  The paper ends in Sect. \ref{summary} where we draw conclusions and summarize the results and the implications of our calculations.

 \section{The binary model and its specific heat}
 \label{Binary}

The model discussed here is the classical example \cite{89DAY,99PH} of a 
glass-forming binary mixture of $N$ particles in a 2-dimensional domain of 
area $V$, interacting via a soft $1/r^{12}$ repulsion with a `diameter' 
ratio of 1.4.  We refer the reader to the extensive work done on this system 
\cite{89DAY,99PH,07ABHIMPS,07HIMPS,07IMPS}. The sum-up of this work is that 
the model is a {\em bona fide} glass-forming liquid meeting all the criteria 
of a glass transition. 

In short, the system consists of an
equimolar mixture of two types of particles, ``large"  with `diameter' 
$\sigma_2=1.4$ and ``small" with 'diameter` $\sigma_1=1$, respectively, but 
with the same mass $m$. In general, the three pairwise
additive interactions are given by the purely repulsive soft-core potentials
\begin{equation}
\phi_{ab}(r) =\epsilon \left(\frac{\sigma_{ab}}{r}\right)^{n} \ , 
\quad a,b=1,2 \ ,
\label{epot}
\end{equation}
where $\sigma_{aa}=\sigma_a$ and $\sigma_{ab}= (\sigma_a+\sigma_b)/2$. The
cutoff radii of
the interaction are set at $4.5\sigma_{ab}$. The units of mass, length, time
and temperature are $m$, $\sigma_1$, $\tau=\sigma_1\sqrt{m/\epsilon}$ and
$\epsilon/k_B$, respectively, with $k_B$ being
Boltzmann's constant. In numerical calculations the stiffness parameter of the
potential (\ref{epot}) was chosen to be $n=12$. 

We turn now to the analysis of the specific heat of this model as a function of the temperature.

\subsection{Specific heat (simulations)}

The specific heat capacity (specific heat) at constant volume is defined by:
\begin{equation}
\frac{C_{V}}{N}=\frac{d}{2}+\frac{\partial}{\partial T} \frac{\langle U\rangle}{N}\Bigg|_{V} \ ,
\label{sphv}
\end{equation}
where $d$ is the space-dimension and the  potential energy  of a binary mixture is given by:
\begin{equation}
 U=\frac{1}{2}\sum\limits_{i\ne j}
\phi_{ab}(r_{ij}) \ . \label{energy}
\end{equation}
Here $r_{ij}$ is the distance between particles $i$ and $j$.
The average value of the potential energy is defined by averaging over
configurational space $\B \Gamma$:
\begin{equation}
\langle U\rangle=\frac{\int U \exp(-\frac{U}{T})\ud \Gamma}
{\int  \exp(-\frac{U}{T})\ud \Gamma} \ .
\label{ustat}
\end{equation}
Substitution of (\ref{ustat}) into (\ref{sphv}) yields the following
expression for the specific heat: 
\begin{equation}
\frac{C_{V}}{N}=\frac{d}{2}+\frac{\langle U^2\rangle-\langle U\rangle^2}{NT^2} \ . 
\label{cv1}
\end{equation}

\begin{figure}[!h]
\centering
\hskip -1 cm
\includegraphics[width= 0.50\textwidth]{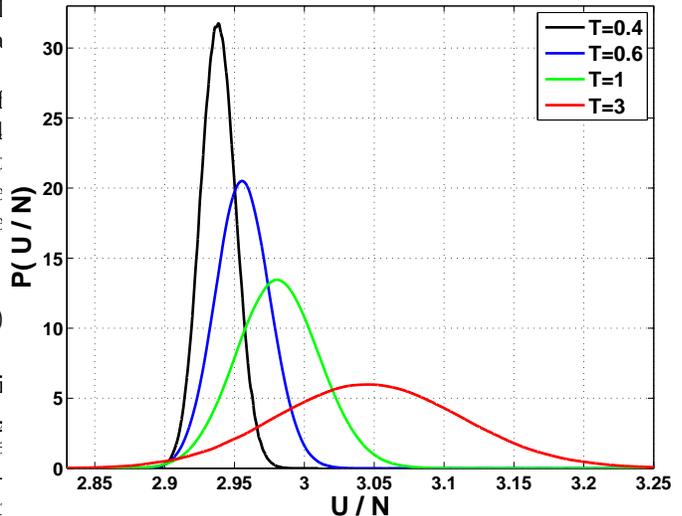}
\caption{Color online: the energy distribution functions for the binary mixture model,
computed at constant volume, such that the volume agrees with the pressure
P=13.5 \cite{99PH} at each temperature.}
\label{Cfig1}
\end{figure}

The specific heat of our binary mixture model was measured at constant volume 
in \cite{87GN,99PH,02PH} and by us. In simulations one can measure the specific heat 
directly from its definition (\ref{sphv}) or (\ref{cv1}). We have used the
last equation which allows one to estimate the specific heat in a single run 
of the canonical ensemble Monte Carlo simulations. At each
temperature the density was chosen in accordance with the simulation results
in an NPT ensemble as described in \cite{99PH} with the pressure value fixed
at $P=13.5$. As the initial configuration in the Monte Carlo process the last
configuration of the molecular dynamics run for this model at given
temperature after $1.3 \times10^8$ time steps was used. After short
equilibration the potential energy distribution functions were measured during
$2\times 10^6$ Monte Carlo sweeps. The acceptance rate was chosen to be $30\%$. 
Simulations were performed with $N=1024$ particles in a square cell with
periodic boundery conditions.

\begin{figure}[!h]
\centering
\hskip -1 cm
\includegraphics[width= 0.50\textwidth]{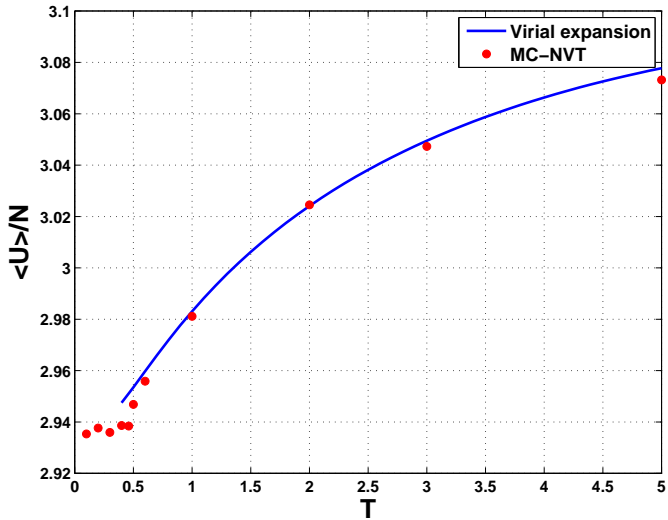}
\caption{Color online: in dots: the temperature dependence of the average potential energy per
in the binary mixture model, computed at constant volume, such that the volume agrees with the pressure P=13.5 \cite{99PH} at each temperature. The continuous line represents the
approximation furnished by the virial expansion, which obviously fails for $T< 0.5.$}
\label{Cfig2}
\end{figure}

\begin{figure}[!h]
\centering
\hskip -1 cm
\includegraphics[width= 0.50\textwidth]{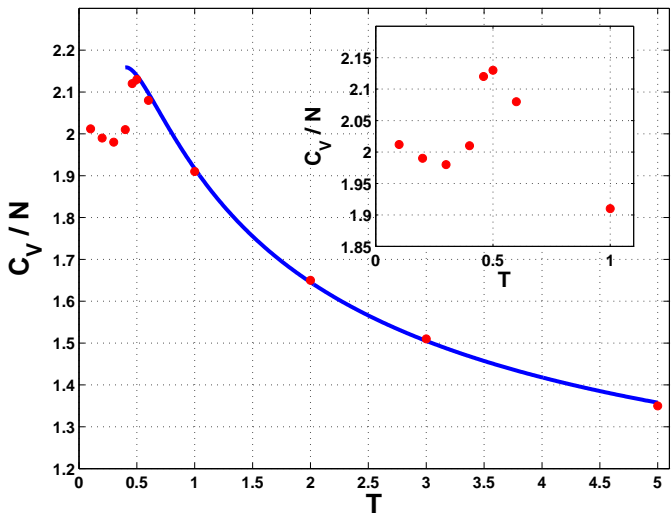}
\caption{Color online: in dots: the temperature dependence of the specific heat in the binary mixture model, computed at constant volume, such that the volume agrees with the pressure P=13.5 \cite{99PH} at each temperature.  The data indicate the existence of two specific heat peaks, one prominent at about $T=0.5$ and a smaller on at about $T=0.1$, and see the inset for finer detail. The continuous line represents the approximation furnished by the virial expansion, which obviously fails for $T< 0.5.$}
\label{Cfig3}
\end{figure}

Examples of the spline interpolation of the potential energy distribution for
a few temperatures are shown in Fig.\ref{Cfig1}. The first and second moments
of these distributions define the average value of the potential energy
(Fig.\ref{Cfig2}) and the specific heat (Fig\ref{Cfig3}). We stress that these results were
computed at constant volume, such that the volume corresponds to simulations in NPT ensemble with the pressure P=13.5 \cite{99PH} at each temperature.

One can see from these figures that the behavior of both quantities, the
first and second moments of the distribution,  change abruptly in the vicinity of 
$T\sim  0.5$. The specific heat displays a maximum in the
temperature dependence.  Our simulations appear to provide trustable values of $C_{V}$
down to lowest temperatures where the value of the specific heat coincides
with that of two-dimensional solid, i.e. $C_{V}=2$. What could not be seen in earlier simulations
is that there is a much smaller second peak
of the specific heat at lower temperatures. To resolve it to the naked eye we present in Fig.
\ref{Cfig3} a blow-up of the region of lowest temperatures where the second peak
is more obvious. To understand the nature of the specific heat anomalies we turn now to a theoretical 
analysis of the caloric equation of state in order to
study the specific heat using the definition (\ref{sphv}). The physical origin of the two peaks will be
explained in Subsect. \ref{physics}. The reader who is mostly interested in the physical insight is invited
to jump to that subsection.

\subsection{Specific heat (series expansions)}
\label{Eqnstate}

The general expression for the pressure (thermal equation of state) obtained 
from the virial theorem is given by:
\begin{equation}P=\rho T-\frac{\rho}{2d}\cdot \frac{1}{N}\langle\sum
\limits_{i\ne j} r_{ij}
\frac{\partial\phi_{ab}(r_{ij})}{\partial r_{ij}}\rangle \ , 
\label{virpres}
\end{equation}
where $\rho$ is the particle number density. The potential is a homogenous function
of degree $-n$ (Eq.~(\ref{epot})), therefore :
\begin{equation}
r\frac{\partial\phi_{ab}(r)}{\partial r}=
-n\phi_{ab}(r) \ . 
\label{der}
\end{equation}
Due to this property of the interaction potential we find a connection between the pressure
and the temperature and mean energy:
\begin{equation}
P=\rho T+ \frac{n}{d}\rho  \frac{\langle U\rangle}{N} \ .
\label{press}
\end{equation}
This equation is exact for one component and multicomponent systems and is 
valid at all temperatures, from liquid to  solid. 

The next simplification for systems with an inverse power inter-particle 
interaction consists in the dependence of all excess thermodynamic properties 
relative to the ideal gas on a single density-temperature variable \cite{70HRJ,71HR}. To see why,
recall that for a one component system the canonical partition function 
$Z_{N}$ is defined by:

\begin{equation}
Z_{N}=\frac{1}{N!\Lambda^{dN}}\int \exp\Bigg(
-\sigma^n\frac{\epsilon}{k_{B}T}\sum\frac{1}{r_{ij}^n}\Bigg) \ud \vec{r}^N. 
\label{e1}
\end{equation}
Here $\Lambda=h/(2\pi mk_{B}T)^{1/2}$ is the thermal de Broglie wavelength.  
The typical distance between particles is given by 
\begin{equation}
l=\Bigg(\frac{V}{N}\Bigg)^{1/d} \ . \label{dist}
\end{equation}
Thus one can use new variables in the integral (\ref{e1}),
$\vec{s}=\vec{r}/l$, and 
the canonical partition function can be rewritten as:
\begin{equation}
Z_{N}=\frac{V^N}{N!\Lambda^{dN}}\frac{1}{N^N}\int \exp\Bigg(
-\rho^{n/d}\frac{\epsilon}{k_{B}T}\sum\frac{1}{s_{ij}^n}\Bigg) \ud\vec{s}^N \
, 
\label{e2}
\end{equation}
where the dimensionless particle number density is defined by
$\rho=\frac{N}{V}\sigma^d$. This way of writing the partition function underlines the
existence of the ideal gas contribution before the configurational integral, and
the dependence of the configurational integral, in the case of one component,  on a natural parameter,  $\Gamma=\rho(\frac{\epsilon}{k_{B}T})^{d/n}$.  

In the case of a multicomponent system the properties of the mixture can be
approximated by those of a one component reference fluid \cite{74H} with an
effective diameter defined by:
\begin{equation}
\sigma_{e}^{d}=\sum\limits_{a,b}x_{a}x_{b}\sigma_{ab}^{d},
\label{diam}
\end{equation}
where $x_{a}=N_{a}/N$ is the particle number concentration. Therefore, the
properties of a mixture are defined by the effective parameter:
\begin{equation}
\Gamma_{e}=\Gamma \Bigg(\frac{\sigma_{e}}{\sigma_{1}}\Bigg)^d.
\label{coup}
\end{equation}

Nevertheless, in \cite{99PH} it was shown, that for soft potential a more suitable
definition of the effective diameter is given by:
\begin{equation}
\sigma_{e}^2=x_{1}\sigma_{1}^2+x_{2}\sigma_{2}^2.
\label{diam1}
\end{equation}
Such a definition leads to a more accurate virial expansion, as
obtained for the present model by \cite{99PH} using molecular dynamics 
simulations in the temperature range $ 0.5\le T\le5$:
\begin{widetext}
\begin{equation}
\frac{P}{\rho T}= 1+1.77306 \Gamma_{e} +2.36241 \Gamma_{e}^2+2.10798 
\Gamma_{e}^3+7.69487 \Gamma_{e}^4
-16.2389 \Gamma_{e}^5+27.99087 \Gamma_{e}^6-16.8643 \Gamma_{e}^7+5.46998 
\Gamma_{e}^8 \ .
\label{presser}
\end{equation}
\end{widetext}
Substitution of the equation (\ref{presser}) to (\ref{press}) yields the
caloric equation of state, the specific heat is calculated after that by 
(\ref{sphv}). The density corresponding to a given temperature is defined as
the solution of the equation (\ref{presser}) at $P=13.5$.

Results of the calculations in the frame of the virial approach are compared
with the simulation results in Figs.~\ref{Cfig2} and \ref{Cfig3}. Clearly, the
virial expansion (\ref{presser}) cannot be trusted for temperatures lower than
$T= 0.5$ since it was computed from simulations that did not go below that
temperature. This temperature is precisely the point at which small micro-clusters
become significant, and a pure liquid homogenous state description of the glassy phase breaks down. Indeed, down to that temperature the prediction of equation
(\ref{sphv}) together with the virial expansion fits the data excellently
well. To understand what happens at lower temperatures we must await Sect. \ref{physics} where
the existence of micro-clusters is taken explicitly into account.
Some comments about the limit of temperature going to zero and the relation to the Madelung
constant can be found in Appendix \ref{madelung}
\subsection{Specific heat (mechanical approach)}
\label{mechanics}
In this subsection we  connect the specific heat to the bulk modulus of the system. 
To this aim  we begin with the microscopic definition of the 
stress-tensor
in an NVT ensemble (see, e.g. \cite{Crox74}):
\begin{equation}
\sigma_{\alpha\beta}=\frac{1}{V}\left(\sum\limits_{i} 
p^{\alpha}_ip^{\beta}_i-\frac{1}{2}\sum\limits_{j\ne i}
\frac{\partial \phi_{ab}(r_{ij})}{\partial r_{ij}}
\frac{r^{\alpha}_{ij} r^{\beta}_{ij}}{r_{ij}}\right),
\label{str}
\end{equation}
where $p^{\alpha}_{i}$ is the $\alpha$ component of the dimensionless momentum
of particle $i$ and $r^{\alpha}_{ij}$ is the $\alpha$ component of the vector 
joining particles $i$ and $j$. The first invariant of the stress tensor is its 
trace:
\begin{equation}
V(\sigma_{xx}+\sigma_{yy})=(\sum\limits_{i}\left(p_{i}^{x}\right)^2
+\sum\limits_{i}\left(p_{i}^{y}\right)^2)+nU. \label{tr1}
\end{equation}
In order to average (\ref{tr1}) one has to take into account that:
\begin{equation}
\langle \sum\limits_{i}\left(p_{i}^{x}\right)^2\rangle=
N\langle \left(p_{i}^{x}\right)^2\rangle=N\cdot T. \label{av1}
\end{equation}
Thus the average of the first invariant (\ref{tr1}) is:
\begin{equation}
(\langle\sigma_{xx}\rangle+\langle\sigma_{yy}\rangle)=2\rho T+
n\rho\frac{\langle U\rangle}{N}. \label{tr2}
\end{equation}
The pressure is defined as $P=(\langle\sigma_{xx}\rangle+\langle\sigma_{yy}
\rangle )/2$ and (\ref{tr2}) yields (\ref{press}). 

The square of (\ref{tr1}) is:
\begin{eqnarray}
&&V^2
(\sigma_{xx}\sigma_{xx}+\sigma_{yy}\sigma_{yy}+2\sigma_{xx}\sigma_{yy})
\label{sq1} \\
&&=\!\!\!\left(\sum\limits_{i}\left(p_{i}^{x}\right)^2
\!+\!\sum\limits_{i}\left(p_{i}^{y}\right)^2\right)^2\!\! \!\!+\!\! 2n\left(\sum\limits_{i}\left(p_{i}^{x}\right)^2
\!+\!\sum\limits_{i}\left(p_{i}^{y}\right)^2\right)U\nonumber\\ &&+n^2U^2.
\nonumber
\end{eqnarray}
To compute the average of this equation we need to use the fact that $\langle
\left(p_{i}^{x}\right)^2\rangle=T$ and 
$\langle \left(p_{i}^{x}\right)^4\rangle=3T^2$.
After averaging (\ref{sq1}) is written as:
\begin{eqnarray}
&V^2&
(\langle\sigma_{xx}\sigma_{xx}\rangle +\langle\sigma_{yy}\sigma_{yy}\rangle +
2\langle\sigma_{xx}\sigma_{yy}\rangle) \nonumber \\
&=&4NT^2+ 4N^2T^2+4NTn\langle U\rangle+n^2\langle U^2\rangle.
\label{av3}
\end{eqnarray}
The square of (\ref{tr2}) is given by:
\begin{eqnarray}
&V^2& (\langle\sigma_{xx}\rangle^2+\langle\sigma_{yy}\rangle^2
+2\langle\sigma_{xx}\rangle\langle\sigma_{yy}\rangle)\nonumber \\
&=&4N^2T^2+4NTn\langle U\rangle+n^2\langle U\rangle^2
\label{sq2}
\end{eqnarray}
After substraction of (\ref{sq2}) from (\ref{av3}) we have:
\begin{eqnarray}
&V^2&
\Bigg[(\langle\sigma_{xx}\sigma_{xx}\rangle-\langle\sigma_{xx}\rangle^2) 
+(\langle\sigma_{yy}\sigma_{yy}\rangle-\langle\sigma_{yy}\rangle^2)\nonumber \\ 
&+&2(\langle\sigma_{xx}\sigma_{yy}\rangle-
\langle\sigma_{xx}\rangle\langle\sigma_{yy}\rangle)\Bigg] \nonumber \\
&=&4NT^2+n^2(\langle U^2\rangle-\langle U\rangle^2).
\label{fl1}
\end{eqnarray}
This equation has a well defined thermodynamic limit since the quantity in square brackets scales like $V^{-1}$. This is seen explicitly in Eq. (\ref{strfluct}).

These results allow us now to find exact relationships between the specific heat, a caloric quantity,
and the elastic constants which are mechanical quantities. To do so we recall the definitions of the elastic constants through
the stress fluctuations (see, e.g., \cite{WP03}):
\begin{eqnarray}
\frac{V}{T}(\langle\sigma_{\alpha\beta}\sigma_{\gamma\delta}\rangle-
\langle\sigma_{\alpha\beta}\rangle\langle\sigma_{\gamma\delta}\rangle)&=&
2\rho T(\delta_{\alpha\gamma}\delta_{\beta\delta}+
\delta_{\alpha\delta}\delta_{\beta\gamma}) \nonumber \\
&+&(C^{B}_{\alpha\beta\gamma\delta}-C_{\alpha\beta\gamma\delta}),
\label{strfluct}
\end{eqnarray}
where $C_{\alpha\beta\gamma\delta}$ are the elastic constants and 
$C^{B}_{\alpha\beta\gamma\delta}$ are the so called Born terms which determine
the instantaneous elastic constants for any given configuration.

Substitution of (\ref{strfluct}) to (\ref{fl1}) yields:
\begin{eqnarray}
& &n^2(\langle U^2\rangle-\langle U\rangle^2)=4NT^2 \nonumber \\
&+&V \Bigg[(C^{B}_{xxxx}+C^{B}_{yyyy}+2C^{B}_{xxyy}) \nonumber \\
&-&(C_{xxxx}+C_{yyyy}+2C_{xxyy})\Bigg]
\label{u1}
\end{eqnarray}

The compression (bulk) modulus in two-dimensional systems is:
\begin{equation}
K=\frac{1}{4}(C_{xxxx}+C_{yyyy}+2C_{xxyy}). \label{k1}
\end{equation}
Recalling Eq. (\ref{cv1}) and substituting it and Eq.~(\ref{k1}) into (\ref{u1}) yields:
\begin{equation}
\frac{C_{V}}{N}=1+4\frac{K^{\infty}-K}{n^2\rho T}. \label{cv2},
\end{equation}
where the bulk modulus in the infinite frequency limit $K^{\infty}=\rho
T+K^{B}$ and 
the Born term is defined by the interpaticle interactions \cite{LM06}:
\begin{equation}
K^B=\frac{1}{4V}\sum\limits_{i\ne j} r_{ij} \Bigg( r_{ij}
\frac{\partial^{2}\phi_{ab}(r_{ij})}{\partial r_{ij}^2}-
\frac{\partial\phi_{ab}(r_{ij})}{\partial r_{ij}}\Bigg). \label{born}
\end{equation}
For the present model the ``infinite frequency" term (cf. \cite{LM06}) is given by:
\begin{equation}
K^\infty=\rho T+\frac{n(n+2)}{4}\rho\frac{\langle U\rangle}{N}. \label{borns}
\end{equation}

This is as much as one can go using exact identities. We reiterate that 
Eq. (\ref{cv2}) is very
interesting, allowing us to connect the bulk modulus to the specific heat. 
In fact, this connection implies that the specific heat measures the difference
between the bulk modulus and its infinite frequency limit. At low temperatures
this difference in the harmonic approximation as given by:
\begin{equation}
K^{\infty}-K=\frac{n^2}{4}\rho T \ ,
\label{moddif}
\end{equation}
independent of the solid structure in contrast to the shear modulus (cf. \cite{07IMPS}).

The bulk modulus $K$ cannot be computed exactly using identities, and we need 
further
information to evaluate it. Fortunately we can estimate the bulk modulus from 
the virial expansion (\ref{virpres}) at $T> 0.5$, since :
\begin{equation}
K=\rho\frac{\partial P}{\partial \rho}. \label{K}
\end{equation}
Having done so we can compare the measurements to what we expect 
theoretically. The specific heat as predicted
from the virial expansion is shown in Fig. \ref{Cfig3} as the blue 
(continuous) line. We should stress that computing $C_V$ from either Eq. (\ref{sphv}) or Eq. (\ref{cv2})
(using the virial expansion (\ref{presser}), yield essentially identical results that cannot be distinguished
in the blue line in Fig. \ref{Cfig3} for $T> 0.5$.

\section{Specific heat - the physical explanation}
\label{physics}
In this section we propose the physical picture behind the existence of the specific heat peaks.
We argue that the specific heat responds to the micro-melting of the clusters - those of small particles at the lowest temperatures and those of the larger particles at higher temperatures. The large increase in
the number of degrees of freedom when a particle leaves a crystalline cluster and joins the liquid background is the basic reason for the increase in entropy that is seen as a specific heat peak.
We can specialize these observations for the model at hand (with inverse power potential)
or present the discussion in greater generality for any model. These two approaches are presented in the two following subsections.

\subsection{Mechanical Equation of State}
In this subsection we employ the mechanical equation of state derived above from which the specific
heat will be computed. To start we define $v_w^\ell$ , $v_w^s$, $v_c^\ell$ and $v_c^s$ respectively as the volume of large particle
in the whey, small particle in the whey, large particle in the solid and small particle in the solid.
Similarly we denote by $ \epsilon_w^\ell$ , $\epsilon_w^s $, $ \epsilon_c^\ell$ and $\epsilon_c^s $ the energy of a large and small particle in the  in the whey and in the crystalline phase respectively.  Needless to say, all these quantities are temperature and pressure dependent; we will therefore explicitly use our low temperature knowledge concerning $v_c^\ell$ and $v_c^s$ in the crystalline phase, but treat the 
difference $v_w^\ell-v_c^\ell$ and $v_w^s-v_c^s$ as constants that we estimate below from our simulation knowledge. Similarly we estimate $\epsilon_c^\ell$ and $\epsilon_c^s$
from our knowledge of the hexagonal lattices at $T=0$. We assume that $\epsilon_w^\ell\approx \epsilon_c^\ell$ and similarly $\epsilon_w^s\approx \epsilon_c^s$ since our simulations indicate
a very small change in these parameters, see Table \ref{table}. It should be stressed that the enthalpy
change at these pressures are almost all due to the $PV$ term.
This will result in a semi-quantitative theory ascribing the important changes in specific heat to the changes in the fraction of particles in curds and whey. In other words the number of particles in the whey and the number of clusters are all explicit functions of temperature and pressure.
\begin{table}[h!]
\centering
\begin{tabular}{|c|c|c|c|c|c|c|c|}
\hline
$\epsilon_c^\ell$&$\epsilon_c^s$&$\epsilon_w^\ell$&$\epsilon_w^s$&$v_c^\ell$&$v_c^s$&
$v_w^\ell$&$v_w^s$\\
\hline
3.69&2.07&3.76&2.16&1.43&0.92&1.58&0.94\\
\hline
\end{tabular}
\caption{Parameters used in the calculation of the specific heat}
\label{table}
\end{table}

As the condensed phase consists of clusters of large and small partilces, we use the notation $N_n^\ell$ for the number of clusters of $n$ large particles and $N_m^s$ for the clusters of $m$ small particles.  In the next subsection we write the energy of our system explicitly in terms of
these cluster numbers. Here however we only need the intensive variables $p_c^\ell=2\sum_nN_n^\ell/N$, $p_c^s=2\sum_mN_m^\ell/N$ $p_w^\ell=2N_w^\ell/N$ and $p_w^s=2N_w^s/N$ which stand for the fraction of large particles and small particles in the curds,  and large particles and 
small particles in the whey, such that $p_c^\ell+p_w^\ell=1$ and $p_c^s+p_w^s=1$. 
Using these variables we can write an expression for the volume per particle $v\equiv V/N$:
\begin{equation}
v=\frac{v_w^\ell+v_w^s}{2} +\frac{v_c^\ell-v_w^\ell}{2}p_c^\ell+\frac{v_c^s-v_w^s}{2}p_c^s \ . \label{vol}
\end{equation}

At this point we need to derive expressions for $p_c^\ell$ and $p_c^s$. To do so we need to 
remember that in the relevant range of temperatures the large particles in the whey can occupy
either hexagonal or heptagonal Voronoi cells, whereas small particles can occupy only pentagonal
or hexagonal cells \cite{07ABHIMPS,07HIMPS,08LP}. Accordingly there are  $g^\ell_w\approx (2^6 -1)/6+2^7/7$ ways to organize the neighbours of a large particle in the whey (neglecting the rare large particle in heptagonal neighbourhood), but only one way in the cluster. Similarly, there are  $g^s_w\approx (2^6-1) /6+2^5/5$ ways to organize a small particle in the whey. We note that this estimate assumes that the relative occurrence of the different Voronoi cells is temperature independent. While reasonable at higher temperatures \cite{08LP}, at lower temperature one should use the full statistical mechanics as presented in
\cite{07HIMPS} to get more accurate estimates of $g^\ell_w$ and $g^s_w$. This is not our aim here; we 
aim at a physical understanding of the specific heat peaks rather than an accurate theory. We thus
end up with the simple estimates
\begin{eqnarray}
p^\ell_c (P,T) 
&\approx& \frac{1}{1+g^\ell_w e^{[(\epsilon_c^\ell-\epsilon_c^\ell)+P (v_c^\ell-v_w^\ell)]/T}} \ , \label{pellc}\\
p^s_c(P,T)& \approx & \frac{1}{1+g^s_w e^{[(\epsilon_c^s-\epsilon_w^s)+P (v_c^s-v_w^s)]/T}} \ . \label{psc}
\end{eqnarray}

It is important to note that the combination of Eq. (\ref{vol}) together 
with Eqs. (\ref{pellc}) and (\ref{psc}) provides a mechanical equation of state that is alternative to the virial expansion presented above. Whereas the latter is best at temperature higher than $T\approx 0.5$
we expect the present one to be best at low temperatures because only the present approach
takes into account the formation of clusters explicitly. The virial expansion by construction is a liquid theory.  We will now compute $C_v$ directly from Eq. (\ref{cv2}). The peaks in the specific heat are determined by the temperature dependence of $p^\ell_c (P,T) $
and $p^s_c (P,T)$ each which has a temperature and pressure derivatives that peaks at a different temperature, denoted as  $T^\ell(P)$ and $T^s(P)$, and see below for details. As said above
we take $\Delta v^\ell\equiv v_w^\ell-v_c^\ell$ and $\Delta v^s \equiv v_w^s-v_c^s$ as approximately constants (as a function of temperature and pressure). The constants are estimated from the condition that the second temperature derivative of 
 $p^\ell_c (P,T) $ and $p^s_c (P,T)$ should vanish. Explicitly:
 \begin{equation}
 \Delta v^\ell\ \approx T^\ell(P^*) \ln g_w^\ell/P^*\ , \quad  \Delta v^s\ \approx T^s(P^*) \ln g_w^s/P^* \ ,
 \end{equation}
where $P^*$ is the pressure for which the peaks in the derivatives are observed (13.5 in our simulations). This is equivalent to a linear dependence of the specific heat peaks as a function
of pressure, $T^\ell(P)/T^\ell(P^*)=P/P^*$ and similarly for the small particles.

In terms of these objects we can rewrite
\begin{eqnarray}
&&v =v_c(P,T)+  \Delta v^\ell(1-p_c^\ell)+ \Delta v^s (1-p_c^s) \ , \label{eqstate1}\\
&&\left(\frac{\partial v}{\partial P}\right)_T =\left(\frac{\partial v_c}{\partial P}\right)_T-  \Delta v^\ell \left(\frac{\partial p_c^\ell}{\partial P}\right)_T-\Delta v^s \left(\frac{\partial p_c^\ell}{\partial P}\right)_T \ .\label{eqstate2}
\end{eqnarray}
To compute the temperature dependence of $\left(\frac{\partial v}{\partial P}\right)_T$ we need first
to determine its $T\to 0$ limit, which is determined by the first term on the RHS of Eq. (\ref{eqstate1}) as the other terms on the RHS decay exponentially fast when $T\to 0$. Since we have already exact
results for the bulk modulus for the present model, we return
to Eqs. (\ref{cv2}) and (\ref{borns}).  We know on the one hand that $\lim_{T\to 0} C_v=2$ and that
$\langle U\rangle /N\approx 2.94$ over the whole interesting temprature range, cf. Fig. \ref{Cfig2}.
The compressibility $\kappa$  is related to the bulk modulus via
$\kappa=-\left(\frac{\partial v}{\partial P}\right)_T/v=1/K$ and therefore easily estimated as $T\to 0$ since there $\left(\frac{\partial v_c}{\partial P}\right)_T\approx -1/(123.5-35T)$. We use this approximation up
to $T\approx 0.5$. 

\begin{figure}
\centering
\hskip -1 cm
\includegraphics[width= 0.55\textwidth]{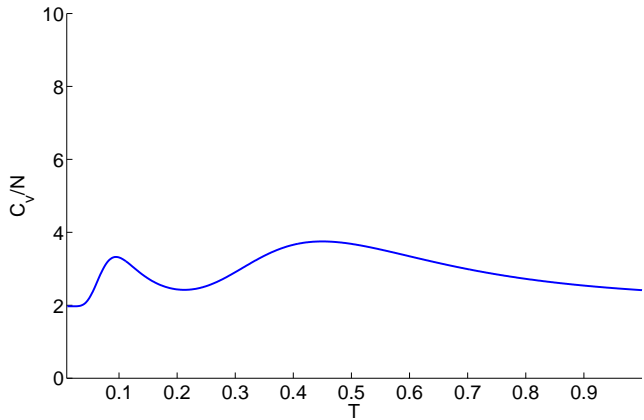}
\caption{Specific heat at constant volume as predicted by the simple theory which is based on
the mechanical equation of state supplied by Eqs. (\ref{ener}),(\ref{vol}) and (\ref{pellc}) and (\ref{psc}). Note that the theory predicts the two peaks which are associated with the micro-melting or micro-freezing of the clusters of large and small particles respectively. The magnitude of the peaks is too high, reflecting terms missing in the simple approach, like the effect of anharmonicity at the lowest temperatures which are negative, tending to decrease the height of the low-temperature peak.}
\label{Cvtheory}
\end{figure}

Having all the ingredients we can compute $C_v/N$ . The parameters used were
estimated from the numerical simulation and are summarized in Table \ref{table}. Since the 
aim of this subsection is only semi-quantitative, we do not make any attempt of parameter fitting, and show the result of the calculation in Fig. \ref{Cvtheory}.

Indeed, the theoretical calculation exhibits the existence of two, rather than one, specific heat
peaks.  We can now explain the origin of the peaks
as resulting from the
derivatives $\left(\frac{\partial p_c^s}{\partial P}\right)_T$ and $\left(\frac{\partial p_c^\ell}{\partial P}\right)_T$. These derivatives change most abruptly when the micro-clusters form (or dissolve), each
at a specific temperature determined by $(h_w^s-h_c^s)/\ln{g_w^s}$ and $(h_w^\ell-h_c^\ell)/\ln{g_w^\ell}$. Note that there can be pressures (both upper and lower boundaries) where the 
the sign of $(h_w^s-h_c^s)$ or $(h_w^\ell-h_c^\ell)$ change sign and the peak can be lost.
\subsection{Caloric Equation of State}

In this subsection we present a more general approach which does not take direct input from results derived for the inverse power potential.  Thus although we use below some parameters read from the simulation, the derivation is very general any pertains to any distribution of clusters. To this aim
we derive a second equation of state, a caloric one. It is quite standard to have two equations of state, only for ideal gas and inverse potentials the two equations of state degenerate into one.

Denote the total energy of the system as a sum of $E_c$, the energy of the clusters (or curds), and $E_w$, the energy of the liquid background (or whey), i.e  :
\begin{equation}
E=E_c+E_w \ .
\end{equation}

These energies are sums over the degrees of freedom - translational, rotational, vibrational and configurational:
\begin{eqnarray}
E_c&=&E_{tr,c}+E_{rot,c}+E_{vib,c}+E_{conf,c} \ , \\
E_w&=&E_{tr,w}+E_{vib,w}+E_{conf,w} \ .
\end{eqnarray}
To estimate $E_{tr,c}$ we consider the number $N_n^\ell$ of clusters of  $n$ large particles and
$N_m^s$ of clusters of $m$ small particles and write
\begin{equation}
E_{tr,c} = T[\sum_n N_n^\ell + \sum_m N_m^s] \ .
\end{equation}
On the other hand in the whey we follow Eyring \cite{69EJ} and Granato \cite{02Gra} and write
\begin{equation}
E_{tr,w} = T[ N_w^\ell +  N_w^s]f \ .
\end{equation}
where $f\equiv 1-V_w^{(s)}/V_w$ is defined as the fraction of free volume
in the liquid phase compared to the equivalent solid crystalline phase. In other words, 
\begin{eqnarray}
V_w&=&N_w^\ell v_w^\ell +N_w^sv_w^s \ . \\
V_w^{(s)}&=&N_w^\ell v_c^\ell +N_w^sv_c^s \ .
\end{eqnarray}

Similarly, we write
\begin{eqnarray}
E_{rot,c} &=& \case{1}{2}T[\sum_n N_n^\ell + \sum_m N_m^s] \ , \\
E_{vib,c} &= &2T[\sum_n N_n^\ell n + \sum_m N_m^s m] \ , \\
E_{vib,w} &= &2T[N_n^\ell +  N_m^s](1-f) \ , \\
E_{conf,c}&=& \epsilon_c^\ell\sum_n N_n^\ell n + \epsilon_c ^s \sum_m N_m^s m  \ , \\
E_{conf,w}&=& \epsilon_w^\ell N_w^\ell + \epsilon_w^s  N_w^s  \ .
\end{eqnarray}

In terms
of these variable we can rewrite
\begin{eqnarray}
\!\!&&E_{tr}= \case{N}{2}T[\case{p_c^\ell}{\langle n\rangle} + \case{p_c^s} {\langle m\rangle}+(p_w^\ell+p_w^s) f] \ , \\
\!\!&&E_{rot,c}=\case{NT}{4}[\case{p_c^\ell}{\langle n\rangle} + \case{p_c^s} {\langle m\rangle}] , \\
\!\!&&E_{vib}=NT[2(1-f) +(p_c^\ell+p_c^s)f] \ ,\\
\!\!&&E_{conf}=\case{N}{2}[ \epsilon_w^\ell+ \epsilon_w^s+( \epsilon_c^\ell- \epsilon_w^\ell)p_c^\ell
+( \epsilon_c^s- \epsilon_w^s)p_c^s] \ .
\end{eqnarray}
Summing up all the contributions we need to pay attention to the order of magnitude of the various
terms. Since we expect the average size of clusters, at the temperatures of interest, to be of the
order of 30 or so, we can neglect safely all the terms that have average size clusters in the denominator.
With this in mind the expression for the energy of the system takes the form
\begin{equation}
\frac{E}{N}= T[2-\frac{2-p_c^\ell-p_c^s}{2} f]+\frac{\epsilon_w^\ell+ \epsilon_w^s}{2}
+\frac{\epsilon_c^\ell- \epsilon_w^\ell}{2} p_c^\ell + \frac{\epsilon_c^s- \epsilon_w^s}{2} p_c^s  \ . \label{ener}
\end{equation}

This is the caloric equation of state that we were after. Using it, we can compute $C_v$ directly from the thermodynamic identity
\begin{equation}
\frac{C_v}{N} = \frac{C_p}{N} - T \frac{v\alpha^2}{\kappa} \ , \label{Cv}
\end{equation} 
where the thermal expansion coefficient is
\begin{eqnarray}
\alpha&\equiv&\left( \frac{1}{v}\frac{\partial v}{\partial T}\right)_P=\label{alpha}\\
&=&\frac{1}{v}[\frac{(v_c^\ell-v_w^\ell)}{2}\left(\frac{\partial p_c^\ell}{\partial T}\right)_P +\frac{(v_c^s-v_w^s)}{2}\left(\frac{\partial p_c^s}{\partial T}\right)_P ] \ , \nonumber
\end{eqnarray}
and the compressibility is
\begin{eqnarray}
\kappa&\equiv & -\left( \frac{1}{v}\frac{\partial v}{\partial P}\right)_T\label{kappa} \\
&=&-\frac{1}{v}[\frac{(v_c^\ell-v_w^\ell)}{2}\left(\frac{\partial p_c^\ell}{\partial P}\right)_T+\frac{(v_c^s-v_w^s)}{2}\left(\frac{\partial p_c^s}{\partial P}\right)_T] \ .\nonumber
\end{eqnarray}

\begin{figure}
\centering
\includegraphics[width= 0.55\textwidth]{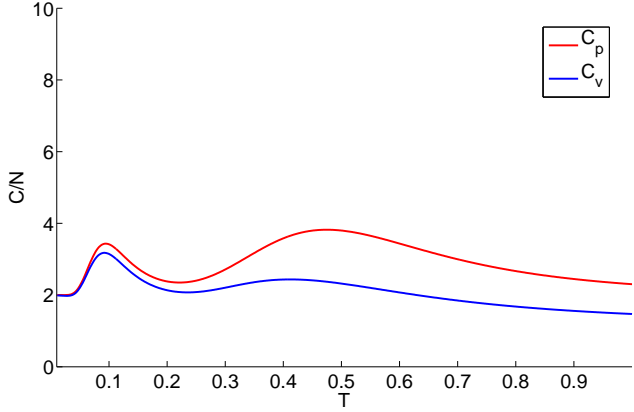}
\caption{Specific heat at constant volume and at constant pressure as predicted by the simple theory which is based on the caloric equation of state supplied by Eqs. (\ref{Cv})-(\ref{Cp}).In agreeement
with the simulations and the theory based on the mechanical equation of state (Cf. Fig. \ref{Cvtheory})
the theory predicts the two peaks to both $C_v$ and $C_p$ which are associated with the micro-melting or micro-freezing of the clusters of large and small particles respectively. Note that $C_p\ge C_v$ as
is expected from thermodynamics.}
\label{Cvtheory2}
\end{figure}

The last object that we need to obtain for evaluating $C_v$ is $C_p$:
\begin{equation}
C_p = \left(\frac{\partial E}{\partial T}\right)_P + P \left(\frac{\partial V}{\partial T}\right)_P \ . \label{Cp}
\end{equation}
Using our expressions (\ref{ener}) and (\ref{vol}) we find
\begin{widetext}
\begin{eqnarray}
\frac{C_p}{N} = \left[2 -\frac{(2-p_c^\ell-p_c^s}{2}f\right]+T\left(\frac{\partial p_c^\ell}{\partial T}\right)_P\left[\frac{f}{2}
+\frac{(h_c^\ell-h_w^\ell)}{2T} \right]
+ T\left(\frac{\partial p_c^s}{\partial T}\right)_P\left[\frac{f}{2}
+\frac{(h_c^s-h_w^s)}{2T} \right]- \frac{(2-p_c^\ell-p_c^s)}{2} T\left(\frac{\partial f}{\partial T}\right)_P \ .
\end{eqnarray}
\end{widetext}
Having all the ingredients we can sum up the terms in Eq. (\ref{Cv}). The parameters used were
estimated from the numerical simulation and are summarized in Table \ref{table}. As before, we did not make any attempt for parameter fitting, and show the result of the calculation in Fig. \ref{Cvtheory2}.

\subsection{Discussion}

The bottom line of the simple theory described in the previous subsection is that there
are two important ranges of temperature, first around $T\approx 0.5$ where clusters of large particles
begin to form, and a second around $T\approx 0.1$ where clusters of small particles begin to appear.
The first important change is also seen in the bulk modulus; this is not surprising, since
the crystalline clusters have a bulk modulus very different from the fluid. Nevertheless between the
clusters we still have appreciable fluid regions which act as lubricants for the response to shear.
We thus expect the shear modulus to change appreciably only when the small particles begin to cluster,
in the vicinity of the smaller specific heat peak. We therefore conjecture that the two specific heat
peaks are also associated with changes in the bulk and shear modulus respectively. We expect that
any  measurements of the glass
properties connected with bulk and shear moduli will show different transition
temperatures if these quantities do not reach simultaneously their $K^\infty$ counterparts.

We cannot at this point asses how general is this split between bulk and shear moduli, and whether
it will be seen in generic glasses. We thus leave this point for further research, stressing that
we expect this phenomenon to appear whenever there exist micro-clusters of preferred oredering
in the scenario of glass-forming.

Having an effective equation of state, albeit approximate, we can easily compute any thermodynamic
derivative of interest. We wrote above explicit expressions for the compressibility and the thermal expansion coefficient. Others are as easily calculated. The point to stress however is how non-universal
the thermodynamics is.  In this model we have two specific heat peaks, in others we might have one
or several. We would also expect a strong pressure dependence for these peaks. 

\section{Frequency-Dependent Specific Heat}
\label{glycerol}

 By applying time dependent heat fluxes $\delta Q(t)$ to the liquid and measuring the resulting temperature fluctuations $\delta T(t)$, the specific heat can be measured $\delta Q(t) = C \delta T(t)$. 
As mentioned in the introduction, measurements of the specific heat of glassy fluids at low temperatures can in principle be made under conditions of either constant volume (isochoric conditions) or constant pressure (isobaric conditions),  but experimentally isobaric conditions are the norm. 

The first and best known measurement of the frequency dependent complex specific heat was performed in glycerol, and we take these experimental results as our motivation for this section.
We stress from the beginning that our approach is not particular for glycerol, and it can be
applied to any other material where, as we assume for glycerol, there exists clusters of
various sizes that determine the dynamical response. In order to develop a model of
the frequency dependent specific heat in glycerol we will employ our own model of the 
 glassy phase of  glycerol. This model assumes that glassy glycerol is a heterogeneous fluid on macroscopic timescales. That is, that while on very long timescales the liquid phase is homogeneous, there exist localized mesoscale domains in the fluid that have macroscopic lifetimes. Indeed, inhomogeneities that appear to survive for $~10^4$ seconds contribute to the dielectric response in the Fourier domain at frequencies as low as $~10^{-4}$ Hz in some cases.  Clearly such imhomogeneities will also contribute to anomalies in the frequency dependent specific heat $C_p(T,\omega )$. We  develop the theory by deriving expressions for the time-dependent enthalpy fluctuations $\langle \Delta H(t)\Delta H(0)\rangle$ that are related to the frequency dependent specific heat at constant pressure in terms of the  distribution of these heterogeneities. The reader is referred to \cite{08HP} for an
 introduction to the dynamical model of glassy glycerol in which the dielectric spectra are computed
 in great detail.

 \subsection{Frequency Dependent Specific Heat}
 
 By considering a temperature field $T(t) = T + \delta T(t), t<0; T(t)=T, t>0$ and using linear response theory  on an isobaric ensemble where the appropriate Boltzmann distribution is $\exp(-\beta H)/Z$, with the enthalpy given by $H=E+PV$, Nielsen and Dyre \cite{96ND} find that the frequency dependent specific heat is given by the form
 \begin{equation}
 \label{linres}
 C_p(T,\omega ) = \frac{\langle \Delta H^2 \rangle}{k_BT^2} + \frac{ i \omega }{k_BT^2}\int_{0}^{\infty}\langle \Delta H(t) \Delta H(0)\rangle e^{+ i \omega t} dt .
 \end{equation}
 In Eq.~(\ref{linres}) $\Delta H(t) = H(t)-\langle H\rangle$ is an enthalpy fluctuation away from equilibrium.
  Therefore we write
 \begin{equation}
 \label{enthalpy}
 H(t) = \sum_s N_s(t) H_s + M_l(t) h_l
 \end{equation}
 where $N_s(t)$ is the number of clusters consisting of $s$ molecules in the glassy phase and $M_l(t)$ are the remaining molecules in the mobile liquid phase. $H_s$ is the enthalpy of a cluster of $s$ molecules at a pressure $P$ and temperature $T$ 
 \begin{equation}
 \label{Hs}
 H_s(P,T) = E_s(P,T) + P V_s(P,T) = (\epsilon_c + p v_c) s + \sigma s^{2/3}.
 \end{equation}
 In Eq.~(\ref{Hs}) $\epsilon_c(P,T)$ is the energy/molecule in the condensed phase; $v_c(P,T)$ is the volume per molecule in the condensed phase; and $\sigma(P,T)$ is the surface energy per molecule. Finally $h_l(P,T)=\epsilon_l+ P v_l(P,T)$ is the enthalpy per molecule in the mobile phase.
 
 Now in equilibrium we can write
 \begin{equation}
 \label{entav}
 \langle H \rangle = \sum_s \langle N_s\rangle H_s + \langle M_l \rangle h_l
 \end{equation}
and we also have the sum rule
\begin{equation}
\label{sum}
\sum_s s N_s(t) + M_l(t) = M
\end{equation}
where $M$ is the total number of molecule in the system.  We can  write the enthalpy fluctuations
away from equilibrium at time $t$ as
\begin{equation}
\label{fluct}
\Delta H(t) = H(t) -\langle H \rangle = \sum_s (N_s(t) - \langle N_s \rangle) h_s = \sum_s \Delta N_s h_s
\end{equation}
where 
\begin{equation}
h_s = H_s - s h_l = (\epsilon_c - \epsilon_l)s + P(v_c - v_l) s + \sigma s^{2/3} \ . \label{hs}
\end{equation}

Let us first calculate the equilibrium fluctuations $\langle (\Delta H )^2\rangle$. To this end
we assume that there are no correlations between the dynamics of clusters of different sizes, implying  $\langle \Delta N_s \Delta N_s'\rangle = 0$. Then,  using the expression Eq.~(\ref{fluct}) for the enthalpy fluctuations, we can immediately write that
\begin{equation}
\label{eqfluct}
\langle (\Delta H )^2\rangle = \sum_s \langle \Delta N_s^2\rangle h_s^2.
\end{equation}
Similarly for the time dependent enthalpy fluctuations
\begin{equation}
\label{tdfluct}
\langle \Delta H(t) \Delta H(0)\rangle = \sum_s \langle \Delta N_s(t) \Delta N_s(0)\rangle h_s^2. 
\end{equation}
We have thus reduced the correlation functions for the enthalpy fluctuations into expressions involving the fluctuations in cluster number for clusters of different sizes $s$. 

To estimate these correlation functions we proceed as follows. First we note that the number of molecules $N_c(t)$  in the clusters can be written as a sum over the clusters as $N_c(t)-\sum_sN_s(t) s$, and consequently the fluctuations in the total number of particles away from equilibrium are 
$\Delta N_c(t) =\sum_s\Delta N_s(t) s$. Then, assuming Gaussian fluctuations we estimate the mean square fluctuations of the number of particles within some small volume
\begin{equation}
\langle (\Delta N_c)^2\rangle \approx \langle N_c\rangle 
\end{equation}
or rewriting 
\begin{equation}
\sum_s \langle (\Delta N_s)^2 \rangle s^2 =\sum_s  \langle N_s\rangle s \ .
\end{equation}
From this equation we therefore see that
 \begin{equation}
 \langle (\Delta N_s)^2 \rangle =\langle N_s \rangle/s \ . 
\end{equation}

For the time dependent fluctuations therefor, assuming an independent Debye relaxation for each cluster, 
\begin{equation}
\label{tdnumfluc}
\langle \Delta N_s(t)\Delta N_s(0)\rangle  = \langle N_s \rangle e^{-t/\tau_s}/s
\end{equation}
where $\tau_s$ is the lifetime of a cluster of size $s$. Thus we get our final expression for the enthalpy fluctuations in equilibrium in terms of the cluster size distribution as
\begin{equation}
\label{eqfluct2}
\langle (\Delta H )^2\rangle = \sum_s \langle N_s\rangle(h_s^2/s). 
\end{equation}
and for their  time dependent correlations
\begin{equation}
\label{tdfluct}
\langle \Delta H(t) \Delta H(0)\rangle = \sum_s \langle N_s \rangle e^{-t/\tau_s} (h_s^2/s). 
\end{equation}

We now substitute these expressions into Eq.~\ref{linres} with the result
\begin{equation}
\label{cp}
C_p(T,\omega ) = \frac{1}{k_B T^2} \sum_s \frac{\langle N_s \rangle (h_s^2/s)}{1 - i \omega \tau_s}
\end{equation}
or splitting the specific heat into its real and imaginary parts
\begin{eqnarray}
\label{reim}
\Re C_p(T,\omega )  & = & \frac{1}{k_B T^2} \sum_s \frac{\langle N_s\rangle (h_s^2/s)}{1 + (\omega \tau_s)^2} \nonumber \\
\Im C_p(T,\omega )  & = & \frac{1}{k_B T^2} \sum_s \frac{\langle N_s\rangle (h_s^2/s)  (\omega \tau_s)  }{1 + (\omega \tau_s)^2} 
\end{eqnarray}

\begin{figure}
\centering
\hskip -1 cm
\includegraphics[width= 0.55\textwidth]{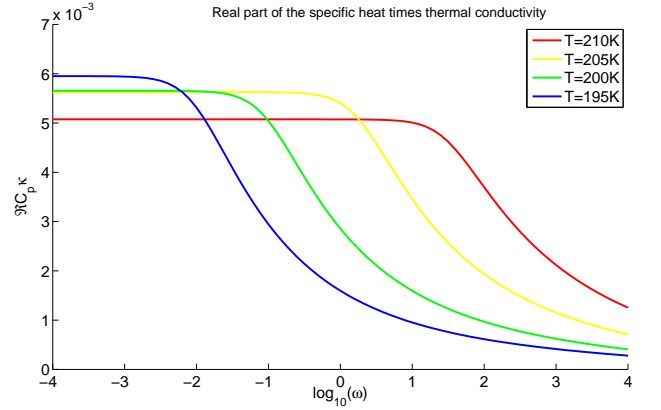}
\caption{The theoretical real part of the specific heat $\Re C_{p}(\omega)$ multiplied by the
thermal conductivity for glycerol.}
\label{fig3}
\end{figure}

\begin{figure}
\centering
\hskip -1 cm
\includegraphics[width= 0.55\textwidth]{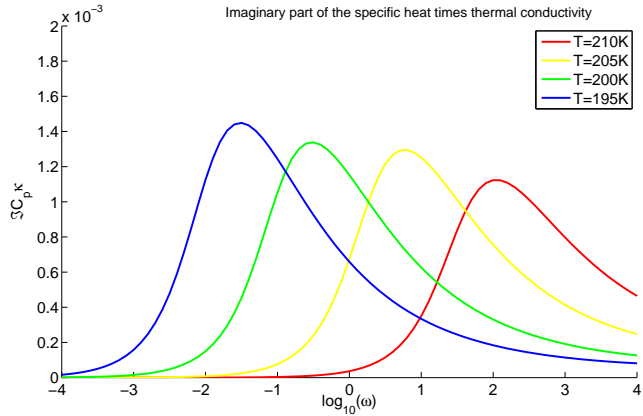}
\caption{The theoretical imaginary part of the specific heat $\Im C_{p}(\omega)$  multiplied by the
thermal conductivity for glycerol.}
\label{fig4}
\end{figure}

We can now use our previous results for the cluster distributions in the case of glycerol to find the real and imaginary specific heat anomalies in the case of glycerol. We do not re-fit any of the parameters used in the calculation of the BDS spectra \cite{09HP}, we simply use the previous knowledge at the temperatures indicated, and plot the results, fitting only the heat conductivity of glycerol. We approximate
$h_s^2/s \approx (h_c-h_\ell)^2 s$, such that Eq  (\ref{cp}) is rewritten as
\begin{equation}
C_p(T,\omega ) \approx  \frac{(h_c-h_\ell)^2}{k_B T^2} \sum_s \frac{\langle N_s \rangle s}{1 - i \omega \tau_s} \ .
\end{equation}
Splitting the specific heat into its real and imaginary parts
\begin{eqnarray}
\Re C_p(T,\omega ) &\approx&  \frac{(h_c-h_\ell)^2}{k_B T^2} \sum_s \frac{\langle N_s \rangle s}{1 + (\omega \tau_s)^2} \ , \\
\Im C_p(T,\omega ) &\approx&  \frac{(h_c-h_\ell)^2}{k_B T^2} \sum_s \frac{\langle N_s \rangle s (\omega \tau_s)}{1 + (\omega \tau_s)^2} \ .
\end{eqnarray}
The resulting curves multiplied by the thermal conductivity are shown in Figs. \ref{fig3} and \ref{fig4}.
These should be compared to Fig.2 of \cite{85BN}. The reader can convince himself that the theory
captures the experimental results quantitatively. 

We would like to stress at this point that the results obtained here are equivalent in dynamical contents
to the computation of the dielectric spectra in \cite{08HP}.  In that calculation one
focused on the dielectric response $\epsilon(\omega)$ and as here decomposed it into its real and imaginary parts $\epsilon (\omega) = \Re  \epsilon (\omega) + i \Im \epsilon (\omega)$. It was
found in \cite{08HP} that without the dc contribution (which is absent in the case of specific heat) we could write
\begin{eqnarray}
\label{result}
\frac{\Re \epsilon (\omega ) - \epsilon_{\infty}}{(\epsilon_0 - \epsilon_{\infty}) }& =&  \frac{\sum_s \langle N_s\rangle  s/[1 + (\omega \tau_s)^2] }{\sum_s  \langle N_s\rangle  s} \\
\frac{\Im \epsilon (\omega ) }{(\epsilon_0 - \epsilon_{\infty}) }& = &  \frac{\sum_s  \langle N_s\rangle  s (\omega \tau_s)/[1 + (\omega \tau_s)^2] }{\sum_s  \langle N_s\rangle  s} \nonumber  . 
\end{eqnarray}
Once normalized the specific heat spectra are identical these spectra . The reason for
the identity is in the assumptions that  $\langle \B m_s\cdot \B m_s\rangle
\sim s$ and $\langle h_s^2\rangle \sim s^2$, cf. Eq. (\ref{hs}). On the other hand
the role of the relaxation times $\tau_s$ and the distribution of cluster sizes are exactly the same
in the two expressions.

\section{Summary and Discussion}
\label{summary}

Probably the most glaring consequence of the calculations presented in this paper is that the specific heat is a valuable indicator of the interesting physics that occurs during the glass transition, but this transition is in no way universal. The temperature dependence of the specific heat is determined by details like inter-particle potentials
and micro-melting or micro-formation of clusters. In this sense any hope for universality is untenable. Nevertheless 
we have shown that the specific heat peaks herald interesting new physics, leading to fast changes in the mechanical moduli which
are also associated with fast changes in the inhomogeneities that are crucial for the glassy behavior, i.e. the formation of micro-clusters. We propose that the appearance of two specific heat peaks in the case of the binary mixture indicates two different ranges for the increase in moduli, the bulk modulus at higher
temperatures when the first type of clusters form, and the shear modulus when the other type of clusters form, and the 'lubricating' effect that allows the system to shear disappears. All this interesting physics is indicated by the behavior of the thermodynamics specific heat. As for the complex specific heat we have shown, in the context
of the example of glycerol, that the physics revealed by the complex specific heat compared to other methods of linear response like Broad Dielectric Spectroscopy are identical. In fact, 
a straightforward consequence of our model for glycerol is the prediction that  the spectra measured from specific heat can be divided by the spectra computed, say, from BDS and the result should be a constant number. We do not have data for exactly the same temperature, but such an experiment would be very useful for the near future.

It is interesting to see in future research whether the two specific heat peaks discussed above may
be seen in other systems, or may be an even richer scenario can appear, with more peaks, when more types of clusters intervene in the process of glass-formation.

\acknowledgments
We are grateful to Sid Nagel and Sasha Voronel for discussion, both in person and electronic,  regarding their measurements of specific heat. This work had been supported in part by the German Israeli Foundation, the Minerva Foundation, Munich, Germany and the Israel Science Foundation.

\appendix
\section{The zero temperature limit}
\label{madelung}

t follows from the simulation results that at lower temperatures the
specific heat is at least close to the value of the solid. Therefore, we can
write the energy of the system in the harmonic approximation:
\begin{equation}
U=U_{0}+\sum\limits_{i,j}^{dN}a_{ij}q_{i}q_{j},
\label{harm}
\end{equation} 
where $U_{0}$ is the potential energy of the system in the reference state, 
$a_{ij}$ are expansion coefficients and $q_{i}$ is a Cartesian coordinate of
the deviation of the current position of a particle from 
equilibrium. There are $dN$ degrees of freedom (neglecting translations and
rotations of the system) in (\ref{harm}), therefore it follows from the
equipartition theorem that the average potential energy of the system in the solid
state is \cite{LL}:
\begin{equation}
\frac{\langle U\rangle}{N}=\frac{U_{0}}{N}+\frac{d}{2}T.
\label{usol}
\end{equation} 
Substituting(\ref{usol}) in (\ref{sphv}) immediately yields $C_{V}=2$
for two dimensional solids.
At the equilibrium configuration the potential energy (\ref{energy}) of the
binary mixture model (\ref{epot}) is given by:
\begin{equation}
U_{0}=\frac{1}{2}\epsilon\sum\limits_{i\ne j}
\left(\frac{\sigma_{ab}}{r_{ij}}\right)^{n}
\label{uref}
\end{equation}
Due to the scaling properties of the inverse power potential it is possible to
normalize the interparticle distances by the typical distance (\ref{dist})
$r_{ij}\to s_{ij}=r_{ij}/l $  \cite{79YH} :
\begin{equation}
\frac{U_{0}}{N}=c_{M}\rho^{\frac{n}{d}},
\label{uref1}
\end{equation} 
where the constant $c_{M}=(1/2N)\sum_{ij}1/s_{ij}^n$ is independent of the
density. Note that this constant is known as the Madelung constant in solid-state physics. Taking into account (\ref{uref1}) the average potential energy of a
harmonic solid (\ref{usol}) can be rewritten as:
\begin{equation}
\frac{\langle U\rangle}{N}=c_{M}\rho^{\frac{n}{d}}+\frac{d}{2}T.
\label{usol1}
\end{equation}
\begin{figure}
\centering
\hskip -1 cm
\includegraphics[width= 0.50\textwidth]{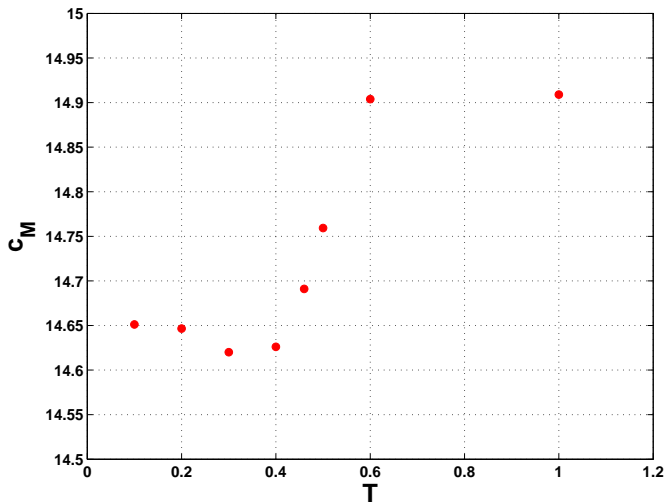}
\caption{Evaluated values of the constant $c_{M}$ from the simulation results.}
\label{Cfig4}
\end{figure}
The value of the constant $c_M$ can be calculated simulationally using Eq. (\ref{usol1}); the results are shown in Fig.\ref{Cfig4}.
One can see that below $T= 0.5$ the change of this constant is small.
Nevertheless,  this change reflects the fact that our calculations at the smallest
values of the temperature are not fully relaxed to equilibrium even though we
took extreme care. Typically at the lowest temperatures the system can be trapped for
incredibly long times in a local minimum of the energy surface, where each local minimum
having slightly different equations of state \cite{07HIMPS}. While we expect the Madelung
constant to be unique for a given crystal, our system here contains clusters of preferred
structures with random orientations \cite{08IPRS}, and therefore the analog of the Madelung constant is
not strictly defined. It may very well depend on the protocol of cooling. The present best
estimate of the value of this parameter at the lowest temperatures is
 $c_M=14.649$.

The caloric equation of state (\ref{usol1}) substituted to the virial equation
(\ref{virpres}) gives the following thermal equation:
\begin{equation}
\frac{P}{\rho T}=(1+\frac{n}{2})+\frac{n}{d}(c_{M}/\sigma_{e})\Gamma_{e}^{n/d}.
\label{eqsol}
\end{equation} 
The value of the renormalized constant $c_{M}/\sigma_{e}=1.394$ can be
compared with the result for the two dimensional one-component system with 
hexagonal crystal, which is $1.268$ \cite{81BGW}. The fact that  this constant is expected
to increase in an amorphous solid was anticipated in \cite{79YH}.

Finally, we note  from (\ref{eqsol}) that in contrast to a crystalline solid the thermal
(caloric) equation of state here remains ambiguous because the value of $c_M$ depends
on the preparation protocol. With this in mind it becomes fruitless to seek the anharmonic
corrections to equation (\ref{eqsol}) as in the case of a one component
system with a well-defined reference state at low temperatures. Nevertheless we stress that the 
specific heat at  constant volume does not suffer from any ambiguity and therefore can be taken good as a good indicator of the solidification.


\end{document}